\documentclass[preprint,prd,tightenlines,nofootinbib,superscriptaddress]{revtex4-1}

\usepackage{color}
\usepackage{graphicx}
\definecolor{red}{rgb}{1,0,0}
\def\lesssim{\ \hbox{\raise 2pt \hbox{$<$} \kern -13pt
                     \lower 3pt \hbox{$\sim$}}\ }
\def\greatersim{\ \hbox{\raise 2pt \hbox{$>$} \kern -13pt
                     \lower 3pt \hbox{$\sim$}}\ }

\def\pythia{{\sc Pythia}}

\def\ptmin{$p_{\textrm{T0}}$}
\def\ptmax{$p_{\textrm{T}}^{\textrm{max}}$}

\input epsf.tex
\def\desepsf(#1 width #2){\epsfxsize=#2 \epsfbox{#1}}

\usepackage{amsmath,bm}
\begin{document}

\hspace*{12.9 cm} {\small DESY 18-003} 

\vspace*{1.4 cm} 

\title{Investigation of the energy dependence of the $p_{\textrm{T0}}$ parameter in the \textsc{Pythia}~8 Monte Carlo event generator.}
\author{P.\ Gunnellini}
\affiliation{Deutsches Elektronen Synchrotron, D-22603 Hamburg}
\affiliation{Institut f\"ur Experimentalphysik , Universit\"at Hamburg, D-22761 Hamburg}
\author{H.\ Jung}
\affiliation{Deutsches Elektronen Synchrotron, D-22603 Hamburg}
\author{R.\ Maharucksit}
\affiliation{Physics Department, Khon Kaen University, Khonkaen, Thailand}

\begin{abstract}
By using predictions from the \pythia~8 Monte Carlo event generator, we determine the energy-dependent turn-over values (\ptmin) of the partonic cross section for the simulation of multiparton interactions. Since the observed energy dependence of the \ptmin\ values is not well described by a power-law function, we introduce an additional energy-dependent term, to better describe the experimental observations in the energy range $\sqrt{s}$ = 0.3-13 TeV. We obtain a similar level of agreement for predictions using various parton densities.         
\end{abstract}

\pacs{}

\maketitle

\section{Introduction} 
In order to simulate a hadron-hadron collision event, standard Monte Carlo (MC) event generators, such as \textsc{pythia}~8~\cite{Sjostrand:2014zea}, are based on a factorized ansatz, for which any hadron-hadron cross section can be written as a product of two non-perturbative process-independent parton density functions (PDF), one for each of the colliding protons, and a perturbative parton-parton cross section. The so-called underlying event (UE) represents the whole additional activity which occurs at lower scales accompanying the hard scattering, and consists of several components, such as initial- and final-state radiation (ISR and FSR, respectively), multiple parton interactions (MPI), and beam-beam remnants (BBR). All the coloured partons produced by these processes are finally rearranged into colourless hadrons, during the hadronization process. Due to all the different contributions, the resulting hadronic collision is a complex multiparticle process. Particularly relevant for the characterization of the UE are the MPI, which consist of numerous additional 2-to-2 parton-parton interactions, occurring within the single collision event. Due to the large increase of the parton density at small longitudinal momentum fractions $x$, the MPI contribution increases with increasing collision energy. \\

The perturbative partonic cross section of a generic process with two incoming and two outgoing partons (the so-called 2$\rightarrow$2 processes), as a function of the exchanged transverse momentum $p_{\text{T}}$, can be expressed as:

\begin{equation}
\frac{\text{d}\hat{\sigma }}{\text{d}p_{\text{T}}^2} \propto \frac{\alpha_{\text{s}}^2(p_{\text{T}}^2)}{p_{\text{T}}^4} ,
\label{diff-cross}
\end{equation}

where $\alpha_{\text{s}}$ is the strong coupling. By integrating the cross section over $p_{\text{T}}^2$, one obtains:
\begin{equation}
\hat{\sigma } \propto \frac{1}{p_{\text{T}}^2},
\label{cross}
\end{equation}

which shows that the total 2$\rightarrow$2 cross section tends to diverge at small values of exchanged transverse momentum. While the simulation of the hard scattering generally involves relatively large $p_{\text{T}}$ values ($p_{\text{T}}$ $>$ 5 GeV), generated MPI processes might reach very low $p_{\text{T}}$ scales ($p_{\text{T}}$ $\sim$ 1 GeV) where the rapid increase of the partonic cross section becomes relevant and might lead to unphysical results. In order to tame the behaviour of the partonic cross section as a function of $p_{\text{T}}$, the \pythia~8 event generator introduces a regularization, by shifting the value of $p_{\text{T}}$ by a quantity \ptmin, leading to a formulation of the partonic cross section as follows:   
\begin{equation}
\frac{\text{d}\hat{\sigma }}{\text{d}p_{\text{T}}^2} \propto \frac{\alpha_{\text{s}}^2\left(p_{\text{T}}^{2}+p_{\text{T0}}^{2}\right)}{\left(p_{\text{T}}^2+p_{\text{T0}}^{2}\right)^2},
\label{diff-cross2}
\end{equation}
which after integration becomes:
\begin{equation}
\hat{\sigma } \propto \frac{1}{p_{\text{T}}^2+p_{\text{T0}}^2}.
\label{cross2}
\end{equation}
Such a cross section does not present any divergence for $p_{\text{T}}$ $\rightarrow$ 0 any longer. In the simulation, \ptmin\ serves as a phenomenological parameter, which can not be obtained from any first principle, but must be determined from data. Many studies have been performed~\cite{Skands:2014pea,Khachatryan:2015pea}, in order to determine the values of \ptmin\ which, as a function of the center-of-mass energy $\sqrt{s}$, is generally between 1.2-2.5 GeV in the $\sqrt{s}$ range of 300-13000 GeV. The \pythia~8 MC event generator implements an energy dependence of the \ptmin\ parameter, according to a power law of the form: 

\begin{equation}\label{endep}
p_{\rm T0}(\sqrt{s})= p_{\rm T0}^{\rm ref} \left(\frac{\sqrt{s}}{\sqrt{s_0}}\right)^\epsilon,
\end{equation}

where $p_{\rm T0}$ is the regularizator of the partonic cross section, which solves its divergent behaviour for $p_{\text{T}}$ $\rightarrow$ 0, $p_{\rm T0}^{\rm ref}$ is the $p_{\rm T0}$ at a reference energy $\sqrt{s_0}$, and the parameter $\epsilon$ determines the energy dependence. This formulation of the \ptmin\ energy extrapolation follows the same energy dependence as the total hadronic cross section. No other options for the \ptmin\ energy dependence are available in \pythia~8. Previous studies have shown the difficulty of describing measurements sensitive to the UE at various center-of-mass energies~\cite{Skands:2014pea,Khachatryan:2015pea} in the range of 300-7000 GeV, measured at different colliders.\\

At a given center-of-mass energy, the amount of simulated MPI in \pythia~8 depends on the turn-over \ptmin, the PDF, and the overlap of the matter distributions of the two colliding hadrons. The MPI processes produce a lot of coloured partons in the final state, creating a dense net of colour lines which spatially overlap with the fields produced by the partons of the hard scattering and with each other. All the generated colour lines may be connected between each other according to the so-called colour reconnection (CR). The CR mechanism implements the possibility for different colour strings to be reconnected and to exchange colour information.\\

In MC event generators, the PDF are a crucial ingredient for the simulation of both the hard scattering and the UE, as they parametrize the distributions of the partons inside each  hadron which cannot be calculated analytically a priori. However, they can be extracted from fits to the data. These fits use analytical calculations of the hard scattering performed at a certain order of the strong coupling $\alpha_S$. From a fit with a leading-order (LO) calculation in $\alpha_S$, one obtains a LO PDF set, with next-to-leading (NLO) or next-to-next-to-leading (NNLO) order calculations, respectively a NLO or NNLO PDF set is obtained. One of the striking differences between LO PDF sets and NLO or NNLO ones is the gluon distribution at small values of longitudinal momentum fractions $x$ and scales $Q^2$, which is rather flat for NLO and NNLO PDF sets while tends to increase quickly for LO PDF sets at small $x$. Note that all PDF sets have a significant
uncertainty in this region, and the ``correct'' behaviour of the gluon distribution at small $x$ is
not yet established. Furthermore, LO PDF can be directly interpreted as parton densities inside the proton and their behaviour can be easily related to measurable quantities, while this is not obvious for NLO or NNLO PDF sets. Because of these reasons, the usage of LO PDF sets for the UE simulation are generally preferred, but nothing prevents to achieve a good description of the measurements, if one uses NLO or NNLO PDF sets.\\

The goodness of the UE simulation provided by MC event generators and corresponding tunes can be tested by comparing predictions with available data. These data are generally measurements of the number of charged particles and their transverse momentum sum in different regions of the phase space relative to the direction of the hardest objects in the event. In particular, the hard object, which might be a jet, a charged particle or a Z boson, identifies a direction in the transverse plane. The transverse plane is then divided into four regions, according to their azimuthal angle: a ``toward'' and an ``away'' region sensitive to the hard scattering and its recoiling object, and two ``transverse'' regions, more sensitive to UE contributions. In recent measurements, the two transverse regions are further divided into separate measurements. The transverse region with the highest activity is called ``transMAX'' while the one with the smallest activity is labelled as ``transMIN''. The charged-particle multiplicity and the transverse momentum sum of the charged particles, measured as a function of the transverse momentum of the leading charged particle, are referred to as ``UE observables'' in the following.\\

In this document, by using UE observables at various collision energies, we determine separately the \ptmin\ values, which best fit the measurements. We use predictions of the \textsc{pythia}~8.226 event generator produced with various PDF sets evaluated at different order in $\alpha_S$. After observing that the obtained \ptmin\ values are not properly predicted by Eq.~\ref{endep}, we introduce a modification of the energy extrapolation by introducing an additional term, according to the formula: 

\begin{equation}\label{endep2}
p_{\text{T0}}(\sqrt{s})= p_{\text{T0}}^{\text{ref}} \,  \left(\frac{\sqrt{s}}{\sqrt{s_0}}\right)^\epsilon + c
\end{equation}
 
where the quantity $c$ is a free energy-independent parameter\footnote{In a preliminary study, the quantity $c$ was considered dependent on the collision energy, in a functional form such as $c$ = $\left(\frac{\sqrt{s}}{\sqrt{s_0}}\right)^{a}$. After performing the fit to the data, it turned out that $c$ is very weakly dependent on the energy, i.e. the quantity $a$ is very close to 0. In order to eliminate any bias on the choice of the reference energy $\sqrt{s_0}$ on $c$, the energy dependence was not further considered.}. The new proposed term of the energy dependence is an attempt of achieving better predictions of \ptmin\ values. However, it does not significantly modify the structure of the power law already implemented in \pythia~8 and the additional term is expected to introduce a little correction with respect to Eq.~\ref{endep}.

\section{Determination of \ptmin\ values at various energies}
In order to determine the best values of \ptmin, fits to observables sensitive to contributions of MPI at soft and semi-hard scales are performed independently at various collision energies. The considered observables are the charged particle multiplicity and average $p_{\text{T}}$ sum densities as a function of the leading charged-particle transverse momentum, $p_{\text{T}}^{max}$ in the transMIN and transMAX regions. Five different sets of measurements of these observables are considered at various collision energies: 300, 900 and 1960 GeV measured by the CDF experiment~\cite{Aaltonen:2015aoa}, 7000 GeV measured by the CMS experiment~\cite{CMS:2012zxa} and 13 TeV measured by the ATLAS experiment~\cite{Aaboud:2017fwp}. 
The region between 0.5 $<$ $p_{\text T}^{max}$ $<$ 1 GeV and the region between 0.5 $<$ $p_{\text T}^{max}$ $<$ 3 GeV are excluded by the fits performed, respectively, at $\sqrt{s}$ = 7 TeV and $\sqrt{s}$ = 13 TeV, since they are found to be affected by contributions of diffractive processes, whose free parameters are not considered in the tuning procedure.\\

Fits are performed for three different PDF sets released by the NNPDF31 collaboration~\cite{Ball:2017nwa}. They refer to the LO, NNPDF31$\_$lo$\_$as$\_$0130, the NLO, NNPDF31$\_$nlo$\_$ as$\_$0118, and NNLO, NNPDF31$\_$nnlo$\_$as$\_$0118, sets. All fits use as baseline the hadronization parameters of the Monash tune~\cite{Skands:2014pea}. Additionally, they use a range for colour reconnection probability equal to 2.17 and an overlap matter distribution modelled by a double gaussian function with radius and fraction matter in the core equal to, respectively, 0.43 and 0.46. These values of parameters were obtained from preliminary tuning attempts using the LO NNPDF31 sets and they were used also for fits with the other PDF fits for consistency. 

The energy reference used for the extrapolation is set to the collision energy of the considered data points used in the fits. This translates into \ptmin\ = $p_{\text{T0}}^{\text{ref}}$ in Eq.~\ref{endep}. This choice is important in order to reduce the number of fitted parameters at each energy and to eliminate any energy dependence in each single fit. The parameters used in the \textsc{pythia}~8 configuration are listed in Table~\ref{tab:parameters}.

\begin{table}[h!]
\centering
\caption{Values of the parameters used in the \textsc{PYTHIA}~8.226 MC event generator related to the overlap matter distribution function and colour reconnection probability. For the \ptmin\ parameter, which is fitted in the tuning procedure, the considered range for the fits is indicated.}
\label{tab:parameters}
\begin{tabular}{l c}
\hline
Tune:pp & 14\\
Tune:ee & 7\\
MultipartonInteractions:pT0Ref        & 1.0 - 3.0 \\
MultipartonInteractions:coreRadius  &  0.43\\
MultipartonInteractions:coreFraction  &  0.46\\
ColourReconnection:range        &  2.17\\
MultipartonInteractions:bProfile & 2\\
\hline
\end{tabular}
\end{table}

Fits to the UE observables are performed by using both the \textsc{Professor 1.4.0}~\cite{Buckley:2009bj} and \textsc{RIVET 2.4.0}~\cite{Buckley:2013} software. About 30 different choices of the \ptmin\ parameter are considered for building the set of anchor points in the one-dimensional parameter space. For each choice of parameters, two million events are generated, so that for each considered bin, the statistical uncertainty of the MC predictions is smaller than the uncertainty of the experimental data. It has been checked that the bin-by-bin envelopes of the different MC predictions encompass well the data points. After running the different predictions, \textsc{Professor} performs an interpolation of the bin values for the considered observables as a function of \ptmin, according to a third-order polynomial function. We checked that the degree of the polynomial used for the interpolation does not influence the tune results. The obtained function $f^{\textrm {\scriptsize b}}(p)$ describes the MC response of each bin $b$ as a function of the vector of the parameters $p$.  The final step is the minimization of the $\chi^2$ function given by the formula:   
\begin{equation}\label{chi2}
\chi^2(p_{\textrm{T0}})=\sum_{O}w_0\sum_{\textrm{\scriptsize b}\in O}\frac{(f^{\textrm{\scriptsize b}}(p_{\textrm{T0}})-R_{\textrm{\scriptsize b}})^2}{\Delta_{\textrm{\scriptsize b}}^2}
\end{equation}
where $R_{\textrm {\scriptsize b}}$ is the data value for each bin $b$ and $\Delta_{\textrm {\scriptsize b}}$ expresses the total bin uncertainty of the data. The experimental uncertainties are assumed to be uncorrelated between data points. The minimization procedure gives in return the values of the parameters which are able to best describe the considered data.

\section{Results} 
The \ptmin\ values obtained from the fits to the data at the various considered energies, as well as the value of the goodness of the fit, are shown in the Tab.~\ref{tab1}.\\

The different PDF sets show quite different values of \ptmin. In particular, the LO PDF set requires more rapidly changing values as a function of energy, than NLO and NNLO PDF sets. This is the impact of the different behaviour of the gluon distribution at small $x$ values, which are more relevant for higher collision energies. In order to reproduce the UE observables, a larger gluon density prefers a smaller amount of MPI contributions (which translates into a larger \ptmin\ value), while larger MPI contributions, i.e. smaller \ptmin\ values, are needed for smaller gluon densities. The \ptmin\ values obtained for the NLO and NNLO PDF sets are very similar to each other. The fact that the gluon distribution at small $x$ is quite flat for the NLO and NNLO PDF sets has the effect that the \ptmin\ values are very weakly dependent on the energy. They range between 1.4 and 1.95 in the energy range of 0.3-13 TeV.

\begin{table}[ht!]
\begin{center}
\caption{The values of the \ptmin\ parameter obtained from the fits to underlying-event observables at the various energies. The uncertainty quoted for each \ptmin\ represents the value obtained in the fit, when allowing an up/down variation of the $\chi^2$, equal to the absolute obtained $\chi^2$. Also indicated for each energy is the goodness of fit divided by the number of degrees of freedom.}
\vspace{5mm}
\label{tab1}
\begin{tabular}{|c| c c | c c | c c|}\hline  
\bf{Energy (TeV)}    &\multicolumn{2}{|c|}{\bf{LO}}          &\multicolumn{2}{c|}{\bf{NLO}}      &\multicolumn{2}{c|}{\bf{NNLO}}\\  &\bf{$\chi^2$/Ndf} & \bf{$p_{\text T0}$} &\bf{$\chi^2$/Ndf} & \bf{$p_{\text T0}$}  &\bf{$\chi^2$/Ndf} & \bf{$p_{\text T0}$}\\ \hline
0.3 &0.60    &$1.54_{-0.02}^{+0.02}$  &0.65    &$1.44_{-0.02}^{+0.02}$    &0.62 & $1.43_{-0.02}^{+0.02}$\\ 
0.9   &0.46    & $1.74_{-0.06}^{+0.02}$  &0.636    &$1.56_{-0.02}^{+0.02}$    &0.71     &$1.52_{-0.02}^{+0.02}$\\
1.96      &0.51    &$1.96_{-0.02}^{+0.02}$   &0.172    &$1.66_{-0.02}^{+0.02}$    &0.53    & $1.63_{-0.03}^{+0.03}$\\
7   &0.68    &$2.35_{-0.03}^{+0.03}$   &0.47   &$1.89_{-0.03}^{+0.04}$   &1.28      & $1.87_{-0.05}^{+0.05}$\\
13   &0.31    &$2.57_{-0.02}^{+0.02}$   &1.18     &$1.96_{-0.03}^{+0.03}$    &1.83      &$1.94_{-0.03}^{+0.03}$\\ \hline
\end{tabular}
\end{center}
\end{table}

\begin{figure}[h]
  \begin{center}
    \includegraphics[width=0.8\textwidth]{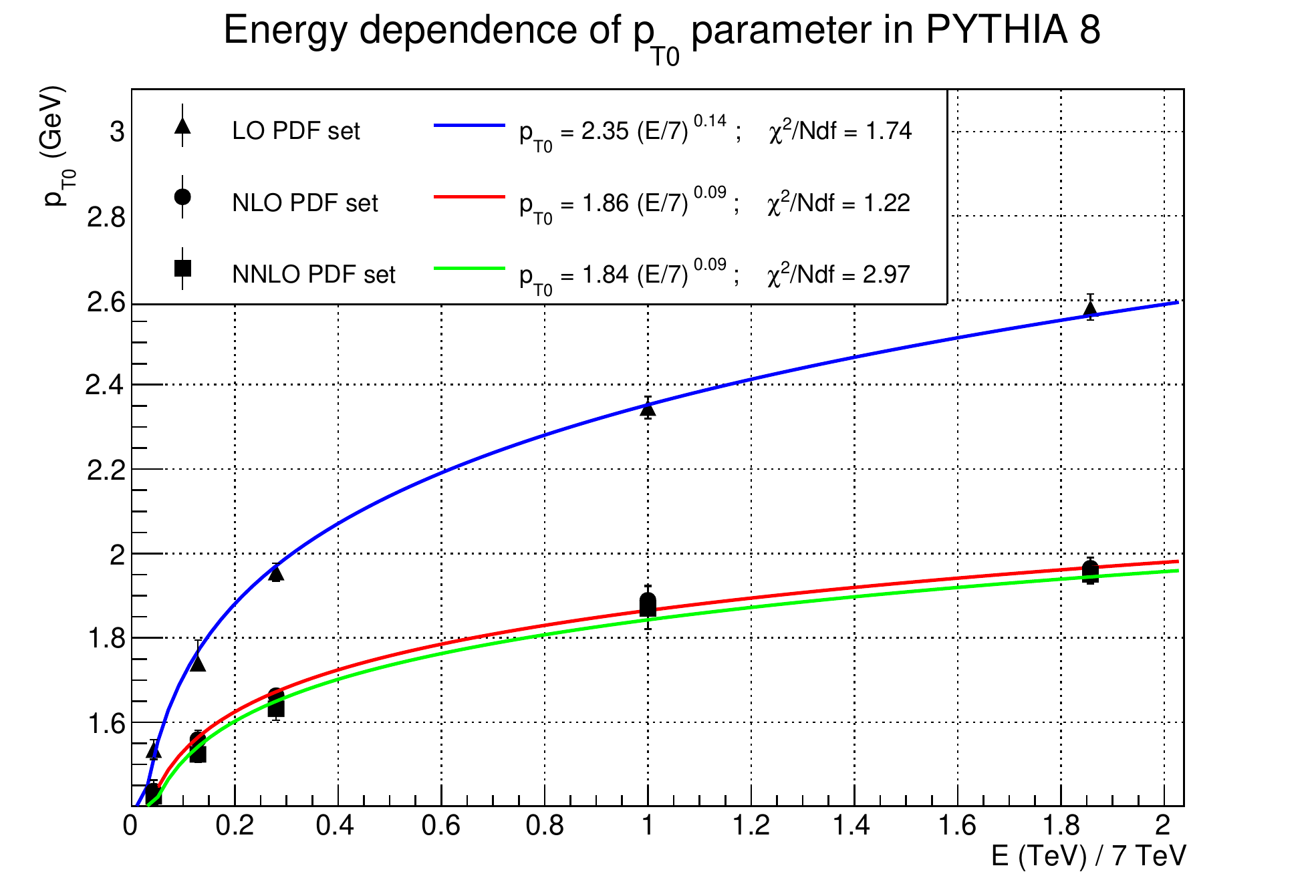}
    \caption{Fit results of the \ptmin\ values according to the energy extrapolation used by default in \textsc{pythia}~8 (Eq.~\ref{endep}) for the three different PDF sets. The values of the obtained parameters as well as the goodness of fit is shown in the plot legend. The energy $E$ is expressed in TeV and an energy reference of 7 TeV is used.}
    \label{Fig:results1}
  \end{center}
\end{figure}

\begin{figure}[h]
  \begin{center}
    \includegraphics[width=0.8\textwidth]{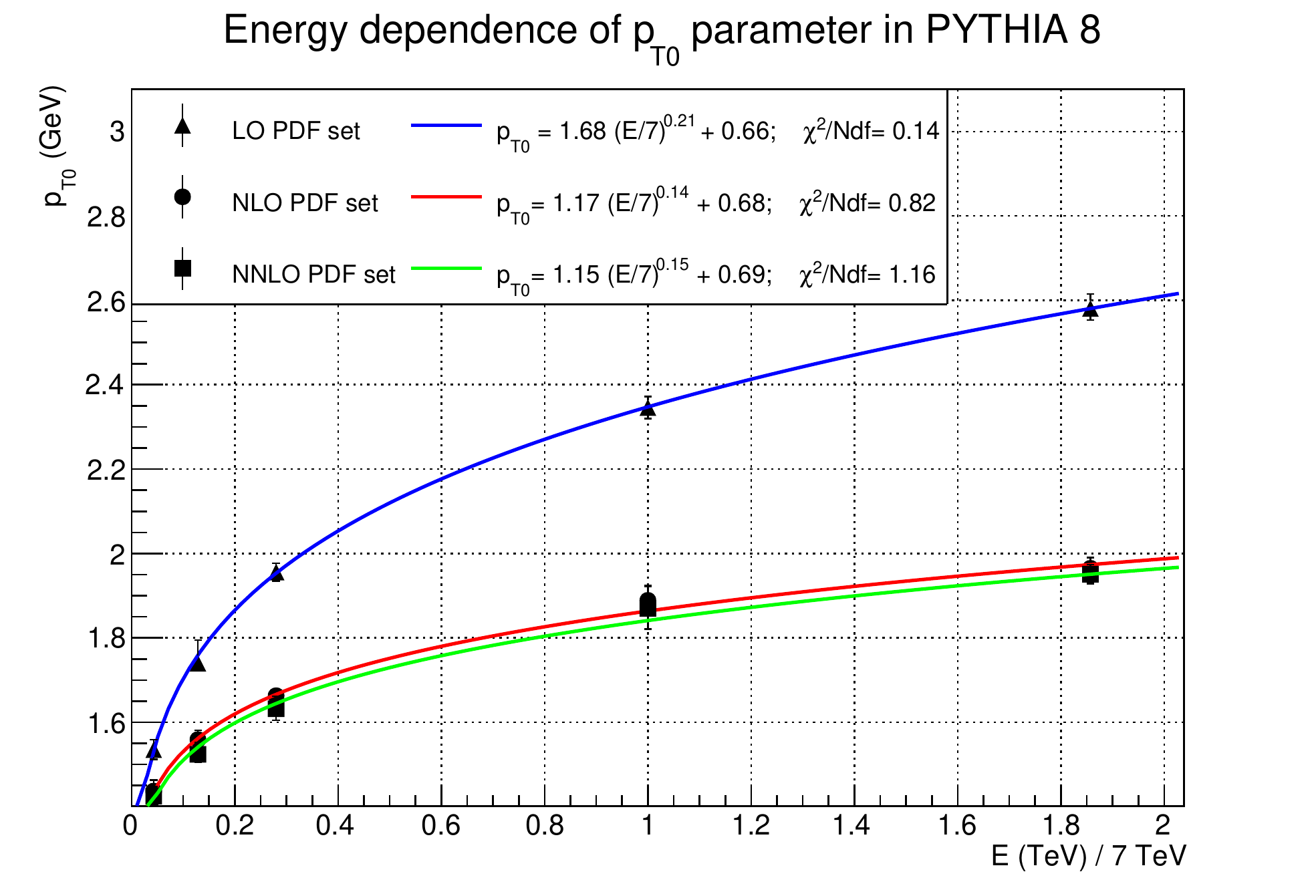}
    \caption{Fit results of the \ptmin\ values according to the energy extrapolation used by default in \textsc{pythia}~8 (Eq.~\ref{endep2}) for the three different PDF sets. The values of the obtained parameters as well as the goodness of fit is shown in the plot legend. The energy $E$ is expressed in TeV and an energy reference of 7 TeV is used.}
    \label{Fig:results2}
  \end{center}
\end{figure}

The results show that for all measurements and considered PDF sets, we are able to obtain a $\chi^2$/Ndf value close to 1. The relative uncertainties obtained for the \ptmin\ values are all of the order of 1-2\%. The \ptmin\ values constitute the input for the fits according to the functions in Eqs.~\ref{endep} and ~\ref{endep2}. The results of the fitting procedure are shown in Figures~\ref{Fig:results1} and ~\ref{Fig:results2} for the two functions, respectively. Table~\ref{tab2} summarizes the parameters of the two functions, as well as the goodness of fit divided by the number of degrees of freedom.   \\

Both functions are able to follow the trend of the \ptmin\ values in the considered energy range. Applying Eq.~\ref{endep}, the low-energy points (300 and 900 GeV), are not well described giving the relatively high $\chi^2$/Ndf values, which go up to 2.97 for the NNLO PDF set. By including the additional term in the energy dependence, the behaviour at low energy significantly improves and the value of $\chi^2$/Ndf decreases down to values close to 1.\\

It has been checked that the modified energy depedence interpolates well, i.e. it gives reliable predictions of \ptmin\ values within the fitted range, for instance, at $\sqrt{s}$ = 2.76 TeV. By comparing \pythia~8 predictions obtained with the \ptmin\ value at $\sqrt{s}$ = 2.76 TeV from Eq.~\ref{endep2} and the parameters of Table~\ref{tab:parameters}, a very good level of agreement is obtained for the UE data measured by CMS at that energy~\cite{Khachatryan:2015jza}.  
The modified energy dependence function can also be reliably used for extrapolation of \ptmin\ values at energies smaller than 300 GeV and higher than 13 TeV and constitutes a alternative to the default \pythia~8 energy extrapolation.

\begin{table}[t]
\begin{center}
\begin{footnotesize}
\renewcommand{\arraystretch}{1.4}
\setlength\tabcolsep{3pt}
\begin{tabular}{|c| r r c |c| r r c |c| r r c |c|}\hline  
\multicolumn{1}{|c}{\bf{Functional form}}    &\multicolumn{4}{|c}{\bf{LO}}   &\multicolumn{4}{c}{\bf{NLO}}      &\multicolumn{4}{c|}{\bf{NNLO}}\\ \cline{2-13} &\multicolumn{1}{c}{\bf{$a$}} &\multicolumn{1}{c}{ \bf{$b$}} &\multicolumn{1}{c}{\bf{$c$}} & \multicolumn{1}{|c|}{\bf{$\chi^2$/Ndf}}   &\multicolumn{1}{c}{\bf{$a$}} & \multicolumn{1}{c}{\bf{$b$}} &\multicolumn{1}{c}{\bf{$c$}} & \multicolumn{1}{|c|}{\bf{$\chi^2$/Ndf}} &\multicolumn{1}{c}{\bf{$a$}} & \multicolumn{1}{c}{\bf{$b$}} &\multicolumn{1}{c}{\bf{$c$}} & \multicolumn{1}{|c|}{\bf{$\chi^2$/Ndf}}\\ \hline
$p_{\text T0}=a(E/7)^b$ & 2.35 & 0.14 & - & 1.74 & 1.86 & 0.09 & - & 1.22 & 1.84 & 0.09 & - &2.97\\
$p_{\text T0}=a(E/7)^{b}+c$ & 1.68 & 0.21 & 0.66  & 0.14 & 1.17 & 0.14 & 0.68 & 0.82 & 1.15 & 0.15 & 0.69 & 1.16 \\ \hline
\end{tabular}
\caption{Summary of the obtained \ptmin\ parameters for the two fitted functions based on LO, NLO and NNLO PDF sets. Also shown is the goodness of fit divided by the number of degrees of freedom. The energy $E$ is expressed in TeV and an energy reference of 7 TeV is used.}
\label{tab2}
\end{footnotesize}
\end{center}
\end{table}

\section{Predictions at $\sqrt{s}$ = 100 TeV}

The new extracted energy dependence can be used to extrapolate \ptmin\ values at high energies. In this Section, we show results for $\sqrt{s}$ = 100 TeV. Note that for the phase space relevant for such an energy, i.e. the gluon distribution at small $x$ values, the current PDF sets are not constrained by any data but only extrapolated from measurements at lower energies. In Table~\ref{tab:100predictions}, the \ptmin\ values for the different PDF sets are listed as predicted by Eq.~\ref{endep}, referred to as "old fit", and by Eq.~\ref{endep2}, referred to as "new fit". While a very small difference is observed between old and new fits for NLO and NNLO PDF sets, the two \ptmin\ values for the LO PDF set differ between each other. This is due to the fact that tunes using a NLO or a NNLO PDF set prefer a very weak \ptmin\ energy dependence, in order to describe measurements at various collision energies. Instead, in tunes using a LO PDF set, one needs a more rapidly increasing \ptmin\ as a function of energy, which is differently predicted by the old and the new fit. 

\begin{table}[h!]
\centering
\caption{Values of \ptmin\ for $\sqrt{s}$ = 100 TeV, as predicted by the old fit (Eq.~\ref{endep}) and the new fit (Eq.~\ref{endep2})}
\label{tab:100predictions}
\begin{tabular}{l | c | c}
\hline
NNPDF order & \ptmin\ from old fit (Eq.~\ref{endep}) & \ptmin\ from new fit (Eq.~\ref{endep2})\\
\hline
LO PDF set & 3.41 GeV & 3.59 GeV\\
NLO PDF set & 2.36 GeV & 2.38 GeV \\
NNLO PDF set & 2.34 GeV & 2.40 GeV\\
\hline
\end{tabular}
\end{table}

\begin{figure}[h]
  \begin{center}
    \includegraphics[width=0.45\textwidth]{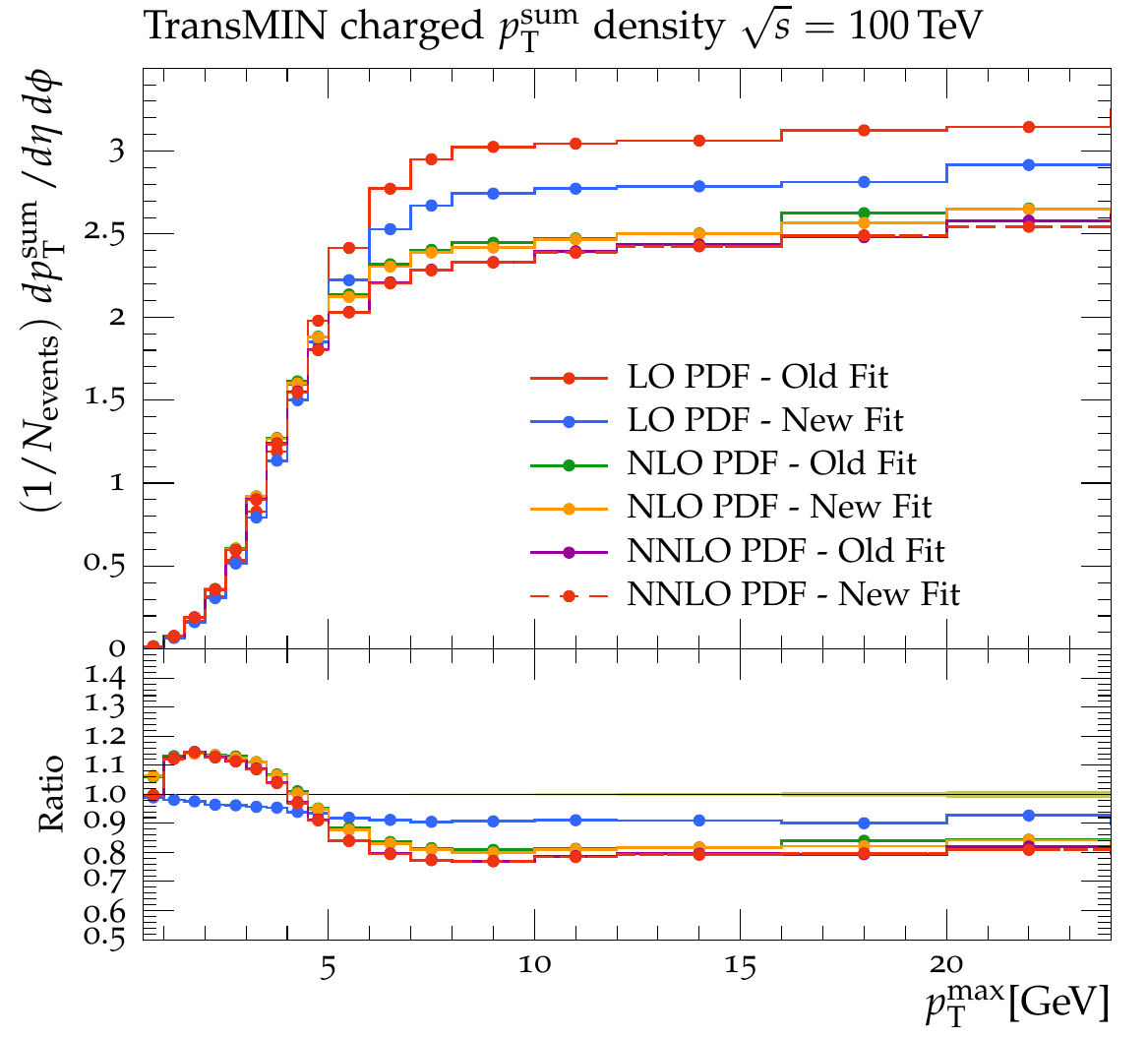}
    \includegraphics[width=0.45\textwidth]{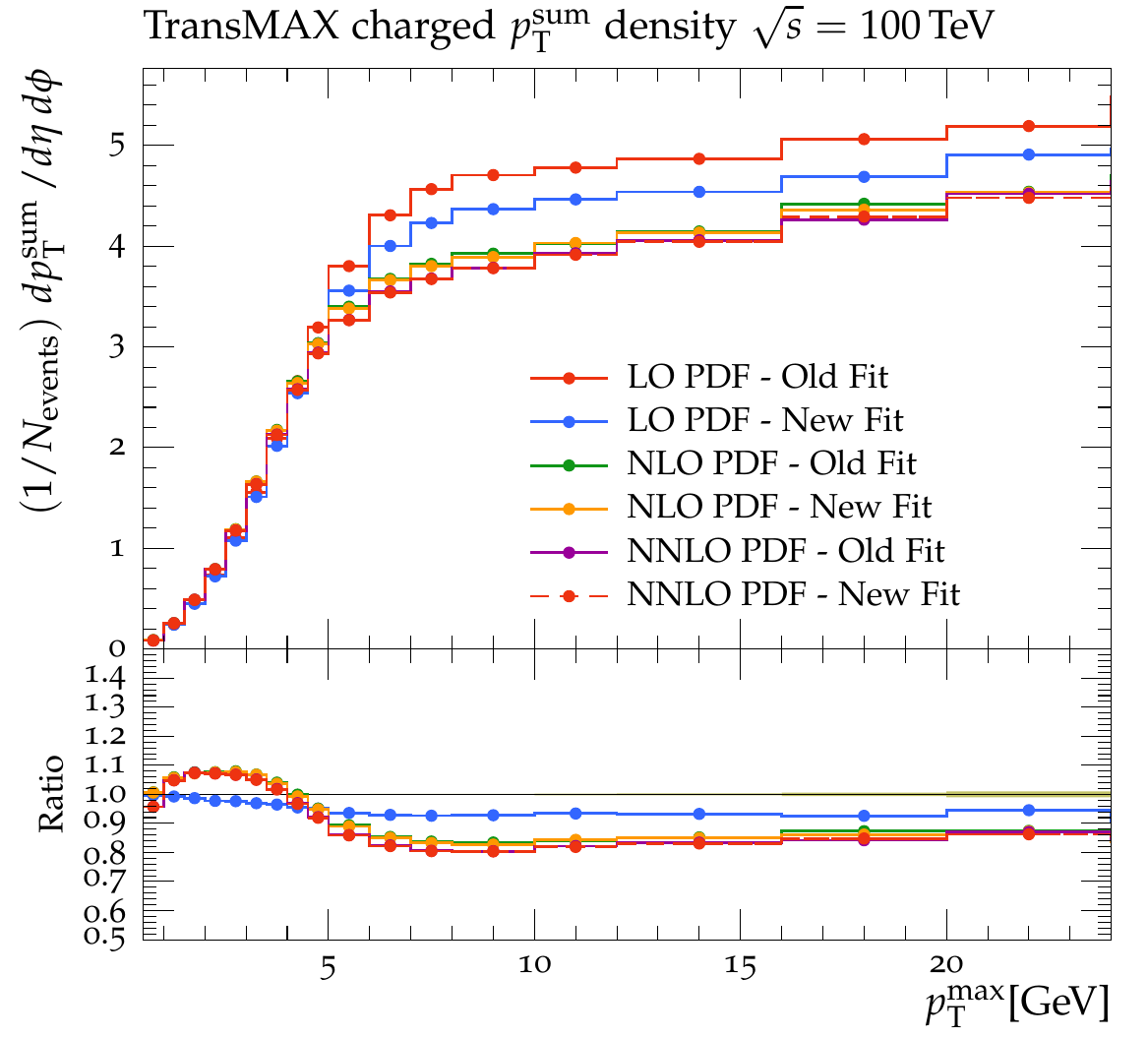}\\
    \includegraphics[width=0.45\textwidth]{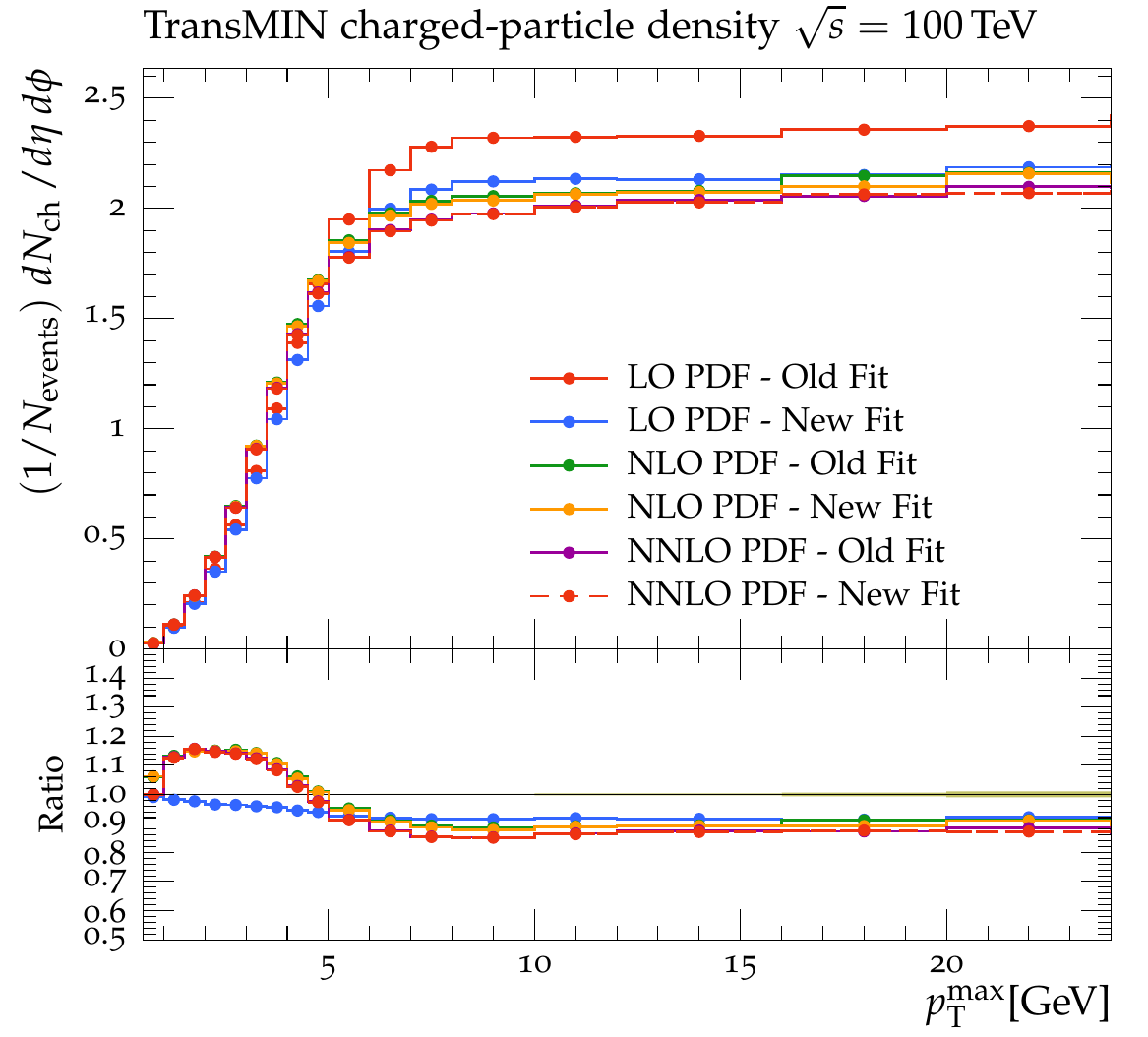}
    \includegraphics[width=0.45\textwidth]{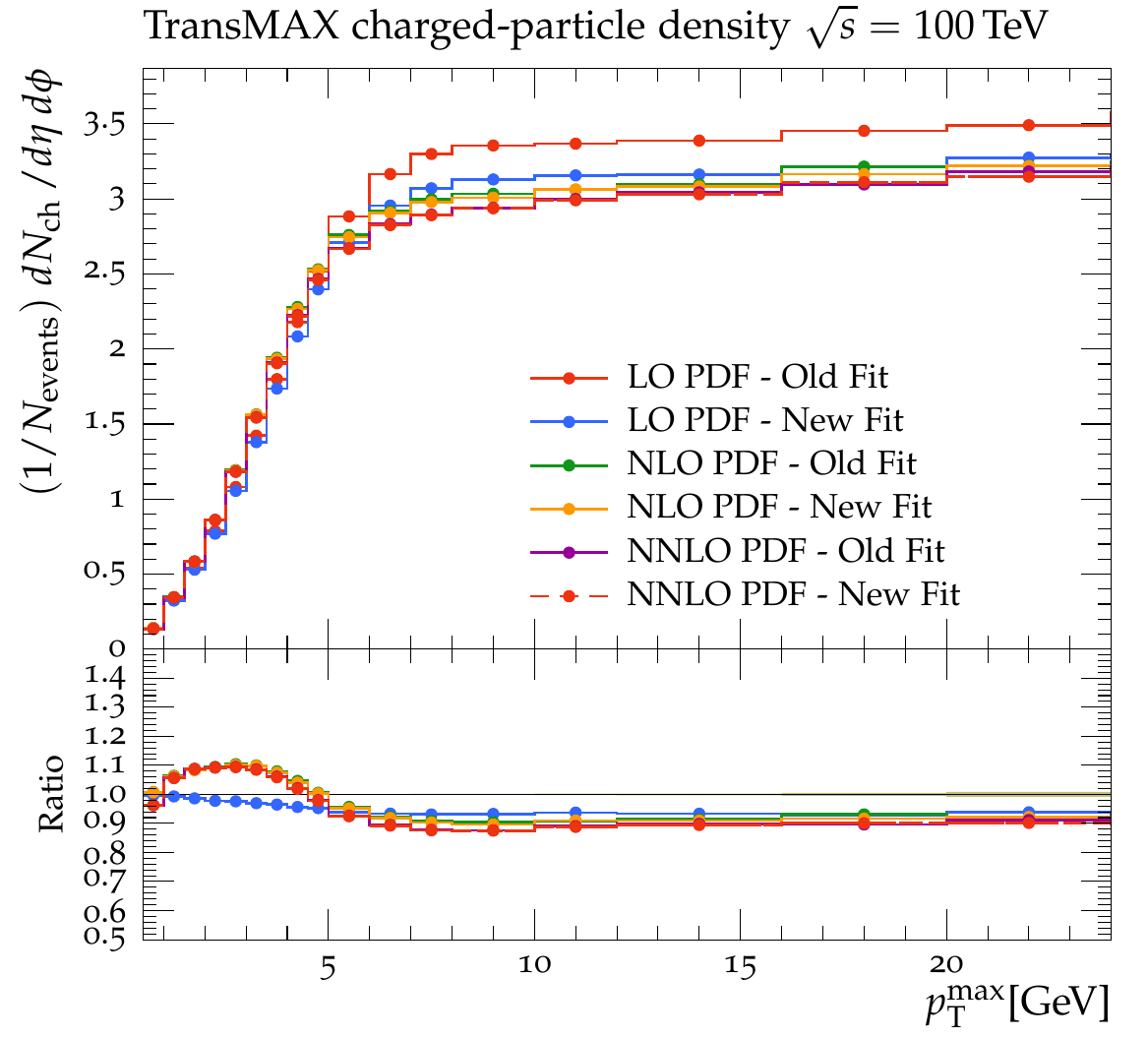}\\
    \caption{Predictions of the \textsc{pythia}~8 tunes at $\sqrt{s}$ = 100 TeV, obtained with the various PDF sets and the \ptmin\ obtained from Eq.~\ref{endep} or Eq.~\ref{endep2}, are shown for average charged particle multiplcities (top plots) and average transverse momentum sum (bottom plots) in the transMIN and transMAX regions, as a function of the transverse momentum of the leading charged particle (\ptmax). Curves labelled as "Old fit" refer to predictions using \ptmin\ as predicted by Eq.~\ref{endep}, while curves labelled as "New fit" use the \ptmin\ value as predicted by Eq.~\ref{endep2} Below each panel, the ratios of all predictions to the ones obtained with the LO PDF set and \ptmin\ from Eq~\ref{endep} are displayed.}
    \label{Fig:plots100}
  \end{center}
\end{figure}

Figure~\ref{Fig:plots100} shows predictions using the various PDF sets on the average charged-particle multiplicity and average charged-particle transverse momentum sum in the transMIN and transMAX regions, as a function of the transverse momentum of the leading charged particle (\ptmax), at $\sqrt{s}$ = 100 TeV. While predictions obtained with the tunes based on NLO and NNLO PDF sets are very similar to each other, independently of the considered energy extrapolation, predictions from the LO tunes differ of up to 10\% between each other. In particular, the new fit predicts a higher \ptmin\ value, and consequently a lower activity in terms of number of charged particles and of transverse momentum. Predictions obtained with NLO and NNLO PDF sets are significantly lower than predictions obtained with LO PDF sets, of less than 10\% if the new fit is used and up to 20\% if the energy extrapolation is carried out through the old fit. 
By performing such measurements at a high collision energy, e.g. 100 TeV, one may be able to validate the performance of the energy extrapolation functions, considered in this document.    

\section{Summary and conclusions}
The energy dependence of the \ptmin\ parameter in the \textsc{pythia}~8 Monte Carlo event generator has been investigated. From observables sensitive to multiparton interactions at low and semi-hard scales at various collision energies, we find that the inclusion of an additional term to the simple power-law function implemented in \textsc{pythia}~8 significantly improves the description of \ptmin\ values obtained at collision energies of 300 and 900 GeV, inferred by measurements performed at the CDF experiment.
This conclusion holds for \pythia~8 predictions using parton distribution functions determined at leading, next-to-leading, or next-to-next-to-leading order in the strong coupling for the underlying event simulation. The additional term is found to be very similar for all parton densities. The modified energy dependence function can be reliably used for extrapolation of \ptmin\ values at energies smaller than 300 GeV and higher than 13 TeV and constitutes a valuable alternative to the default \pythia~8 energy extrapolation.                  

\section*{Acknowledgements}
We thank T. Sj\"{o}strand and P.Skands for comments and suggestions on text and procedure.

\end{document}